# Between Puppet and Actor: Reframing Authorship in this Age of AI Agents

**Yuqian Sun**
Computer Science Research Centre, Royal College of Art
Rausing Research & Innovation Building 15 Parkgate Road
London SW11 4NL UK
yuqiansun@network.rca.ac.uk
ORCID: https://orcid.org/0000-0002-4076-8140

**Stefano Gualeni**
Institute of Digital Games (University of Malta)
Triq L-Esperanto, 20
MSD2080 Msida - Malta
stefano.gualeni@um.edu.mt
ORCID: https://orcid.org/my-orcid?orcid=0000-0003-4873-4751

**Abstract:**
This chapter examines the conceptual tensions in understanding artificial intelligence (AI) agents' role in creative processes, particularly focusing on Large Language Models (LLMs). Building upon Schmidt's 1954 categorization of human-technology relationships and the classical definition of "author," this chapter proposes to understand AI agency as existing somewhere between that of an inanimate puppet and a performing actor. While AI agents demonstrate a degree of creative autonomy, including the ability to improvise and construct complex narrative content in interactive storytelling, they cannot be considered authors in the classical sense of the term. This chapter thus suggests that AI agents exist in a dynamic state between human-controlled puppets and semi-autonomous actors. This conceptual positioning reflects how AI agents, while they can certainly contribute to creative work, remain bound to human direction. We also argue that existing conceptual frames concerning authorship should evolve and adapt to capture these new relationships.

—

The first systematic attempt to categorize the various relationships that human beings can establish with technologies was proposed by German scientist and engineer Hermann Schmidt in 1954. In his essay "Die Entwicklung der Technik als Phase der Wandlung des Menschen," ("Technological Development as a Phase of the Transformation of Man"), Schmidt identifies three stages of technological development, ranking them according to their complexity and their degree of functional autonomy.

1. The first stage is that of **tools**. For Schmidt, tools are simple artifacts that are intuitive in their use and that constitute a direct extension of our native capabilities both physical (such as a hammer or a pair of glasses) or mental (as in the case of an abacus or a notebook). Tools are, in other words, instruments that are actively employed by users to perform specific functions.

2. The second stage of technological development proposed by Schmidt is that of **machines**. These are still relatively simple artifacts, in the sense that we fully understand how they function and can accurately predict their behavior (think of a water mill, for example). When compared with tools, however, machines are characterized by a higher degree of autonomy. A machine can work independently, following a certain plan of action, but still needs to be operated and controlled by its users.

3. The third and last stage in the development of technologies is that of the **automata**. According to Schmidt, automata are **self-regulating artifacts.** They are semi-autonomous technologies in the sense that they do not necessitate operators. Even without the need for supervision during their functioning, automata still require human input, for example when it comes to their design, maintenance, and upgrades. When talking about automata, Schmidt referred to the early products of cybernetic engineering, a field that he contributed to pioneer in the first decades of the twentieth century.

This initial way of understanding our relationships with various types of technology shares the technocratic and utopian tone of many of Schmidt's texts. Nonetheless, Schmidt's original imprint had a lasting impact on how we understand ourselves and our actions within a progressively more technically mediated lifeworld. His original intuitions resonate in a variety of recent frameworks proposed and discussed, among others, by Ihde (1990), Verbeek (2005), or Coeckelbergh (2023) just to name a few.

What about fully autonomous technologies? Authors of science fiction began exploring the idea that technology can be entirely independent of humans in the same period when Schmidt published his famous essay. Philip K. Dick's (2000) short stories such as "Second Variety" and "Autofac" (respectively published in 1953 and 1955) are exemplary in that regard. The prospect of fully autonomous and self-regulating technology still constitutes a fruitful and stimulating scenario for speculative efforts in fields like technoethics and the philosophy of technology (see Coeckelbergh 2010, Gunkel 2018, Gualeni 2020). Conjectures about future technologies might not be directly relevant to the understanding of our present socio-technical context, but they can assist us in addressing the specific themes and questions in scholarly fields like the philosophy of technology or technoethics.

In this text, we focus specifically on how the advances in artificial intelligence (AI) are already impacting our creative agency, especially with regard to the classical notion of authorship. More specifically we discuss what our creative roles and possibilities are when working together with increasingly versatile and semi-autonomous generative AI models such as ChatGPT or Dall-E[1]. Building upon the philosophy of technology and the scholarly field of aesthetics, we aspire to contribute to the understanding of the kinds of technological relationships that we establish with semi-autonomous s. We will pursue those aspirations by initially focusing on the notion of creative agency and by leveraging examples and observations derived from our direct experience working on digital art projects that involve the current generation of generative AI agents.

**On Creative Agency**

In a Reuters article on AI-written books dated February the 21st 2023, Greg Bensinger reports that – at the time of his writing – over 200 e-books in Amazon's Kindle store listing ChatGPT as an author or co-author. The list included titles such as "How to Write and Create Content Using ChatGPT," "The Power of Homework," and the poetry collection "Echoes of the Universe." (Bensinger 2023). In the same article, a person responsible for putting some of those books up for sale explains how, with ChatGPT, he created a 119-page novella about alien factions in a far-off galaxy warring over a human-staffed brothel titled *Galactic Pimp: Vol. 1* in less than a day. The book is available for one dollar on Amazon's Kindle e-book store (ibid.).

As made abundantly clear by Bensinger's article among many others, technologies that can have this kind of creative input and autonomy are already part of our lifeworld. To some, they are an established, familiar component of their creative and professional life. As a case in point, some of the chapters in this very book have been co-written and/or edited with the aid of an artificially intelligent agent trained on a large language model (and so was the one you're presently reading). Our increasingly more familiar relationships with this kind of technology, as anticipated in the introduction section, challenge us to transcend and reframe the traditional understanding of notions related to creative agency such as those of 'author' and 'artwork.'

Why are we focusing specifically on creative agency? As already mentioned, we believe that examining those ideas in the age of semi-autonomous technologies reveals the obsolescence of our conceptual frameworks when thinking about semi-autonomous technologies in ways that are particularly clear.

Let us demonstrate what we mean by introducing what is classically considered to be the author of a creative work. The word 'author' has its etymological roots in the Latin verb *augere*, meaning 'to develop,' 'to grow,' or 'to originate something'. In line with its semantic origins, the author of a work is commonly identified as the individual responsible for its creation. The romantic idea of the tortured genius is a particularly dramatic manifestation of this understanding of authorship (see Hick 2017, 51-53). Art theories that assign creative responsibilities to a special individual who employs their talents and sensitivity to produce something deemed of artistic merit also rely on the understanding of authorship outlined above (see Collingwood 1958; Zangwill 2007, 47-49). The terms 'author' and 'authorship' are still commonly used with that denotation - for example when referring to the people whose names are on the covers of novels, on posters of theater plays, or on the title page of academic texts much like this one. The idea that someone can simply and unproblematically be considered the origin or the cause of a particular event or creation has, however, been put into question long before the emergence of generative artificial intelligences. For a few decades now, a variety of scholars and currents have opposed and criticized this naïve vision of authorship and responsibility, leveraging ideas such as 'actor-networks', 'technological intentionality', 'distributed agency' and 'collective co-creatorship'.[2]

---

[1] In this note, we want to clarify in which way technologies like ChatGPT can be considered semi-autonomous. AI broadly encompasses technologies mimicking human intelligence, with generative AI, like ChatGPT and Dall-E, specializing in producing new content. AI agents are systems designed for autonomous action within certain parameters. They operate within the confines of human programming and the data they have been trained on. Hence, we categorize GPT and similar systems as semi-autonomous, highlighting their ability for significant independent operation yet their reliance on human-created frameworks and inputs.

Going back to the traditional figure of the 'author', what activities and dispositions are necessary for someone to be recognized as such? In fields like the philosophy of art and aesthetics, the classical idea of authorship that was outlined above is generally understood as meeting two conditions:

1) that someone is engaged in the deliberate production of certain aesthetic effects, and
2) their participation in the creative processes that are meant to produce those effects is voluntary (see Anscomb 2022).

Regardless of how loosely one decides to interpret the terms, to be recognized as an author (or a co-author) an agent needs to be able to make conscious choices about what is expressed and elicited by a creative work (1). This way of framing authorship implies the author's capability to deliberately develop and communicate aesthetic decisions (see, among others, Zangwill 2007, 48). According to this perspective, an author thus needs not only to be conscious, but also capable of experiencing and expressing feelings and emotions. To go back to the questions motivating this chapter, it is clear that generative, artificially intelligent agents – at their current state of technological development – cannot be considered either conscious or sentient. Seen through the theoretical frameworks presented above, their incapability to intentionally develop and communicate aesthetic ideas and decisions should bar us from recognizing them as authors or co-authors.

---

[2] It might be worth pointing out that the understanding of creative responsibility implied by this approach applies indifferently to the production of artefacts as well as to performances (and to more ephemeral creations in general). It is in this sense that actors could be considered as co-authoring a theatre play or a movie scene through their particular interpretation of the role they were tasked to perform.

If not authors, could we perhaps consider AI agents to be artificially intelligent co-creators? What we are trying to say is that it would be intuitive to consider generative AI agents as our creative partners (more on this later). After all, the etymology of 'co-creator' refers to someone with whom we work or contribute to crafting something. Having said that, we also need to acknowledge that, in everyday language, those terms are typically used for people, and not when referring to tools, machines, or non-human animals. In addition, both the ideas of authorship and creation connote a voluntary, uncoerced participation in bringing something about (2). As the current generation of AI agents that are not conscious and cannot autonomously decide to participate in the making of something, we should not technically consider them as our creative partners.

We could, of course, always decide to broaden and redefine terms like 'author' and 'co-creator' to fit new ways of approaching creativity and the production of creative works. In other words, we could stretch those ideas beyond their classical and classically anthropocentric origins and in ways that are perhaps less preoccupied with the expression of inner desires and emotions. Alternatively, we could take a different route and propose new frames of reference and new analogies that are better suited to observe and understand our relationship with the present generation of AI agents (and, potentially, future ones as well). The rest of this chapter will demonstrate how AI agents have introduced new complexities to creative work, and why current terminology and analogies are limited in capturing that.

**On AI agents**

The notion of 'agent' has its philosophical roots stretching back to thinkers such as Aristotle and David Hume (Xi et al., 2023). In the broadest sense, an agent is a being with the capability to act, where 'agency' indicates the exercise or manifestation of this capability (Schlosser, 2019). Philosophical discourse traditionally associates agency with intentional actions, driven by desires, beliefs, intentions, as well as the capacity for such actions. In computer science, the term 'agent' refers to a software entity that performs actions on behalf of a user or another program with some degree of autonomy. This definition aligns closely with Schmidt's understanding of automata as "self-regulating artifacts," as those pieces of software can make some decisions and take certain kinds of actions without human intervention.

In November 2023, OpenAI introduced GPTs, enabling users to customize AI agents through straightforward text prompts without coding. This innovation lowered the barrier to designing agents, democratizing the creation process and shifting the discourse around authorship in the digital realm. On platforms like character.ai (Figure 1), users craft agents ranging from adaptations of canonical figures, such as Link from the videogame series *The Legend of Zelda*, to original entities. This burgeoning community, now exceeding 750k on platforms like Reddit, reflects a shift toward more accessible agent creation, as seen in OpenAI's GPT Store where Shakespeare can be reimagined as a nutritionist by inputting a few lines of text as a prompt.

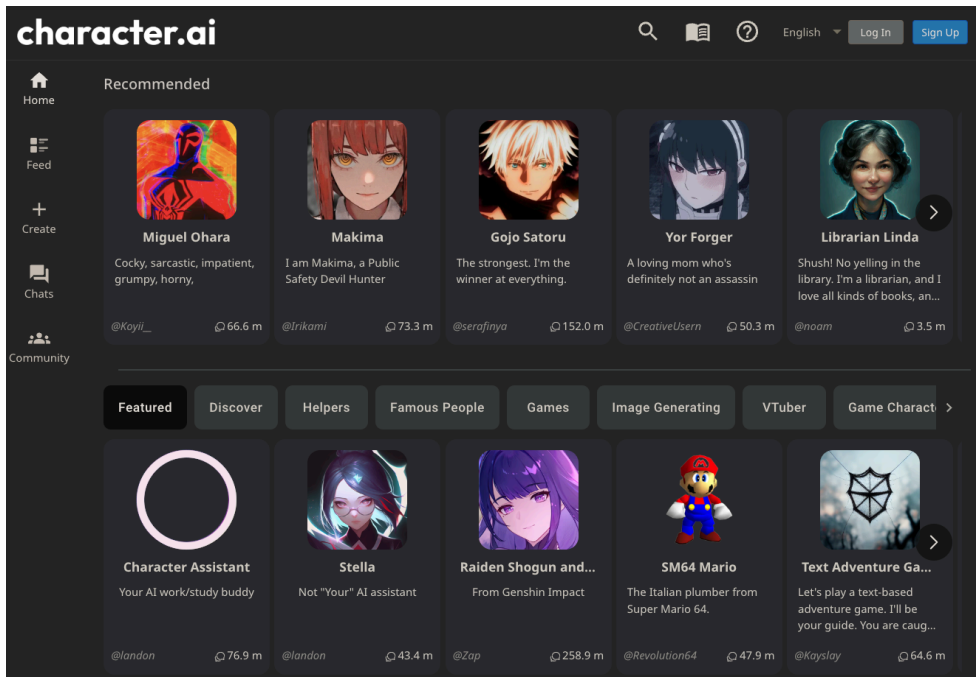

Figure 1: A screenshot of character.ai, a chat application powered by a large language model. Users can easily customize a character chatbot through several lines of descriptions.

As Large Language Models (LLMs) assume the role of agents in creative processes, they also reveal inadequacies and ambiguities in the idea of authorship, classically understood. We suggest that the interplay of metaphors such as 'actor' and 'puppet' might be more fitting when trying to capture our current, creative relationships with LLMs.

The concept of 'puppet,' as puppeteer Bil Baird defines it, is "an inanimate figure made to move by human effort before an audience" (Baird, 1965). Penny Francis describes it as "a representation and distillation of a character, the repository of a persona perceived by both creator and spectator within its outward form" (Francis, 2012). This metaphor emphasizes strong authorial control and limited autonomy - much like a puppet master maintains complete control over their puppet's movements and expressions.

The 'actor' metaphor, in contrast, suggests a higher degree of autonomy and pro-active participation. Like a theater actor who brings their own interpretation to a role while following a director's vision, an AI agent with actor-like qualities can exhibit initiative and unexpected creative behaviors within its programmed parameters. When audiences are immersed in an actor's performance, they might temporarily forget the director's role - similarly, users engaging with sophisticated AI agents might momentarily overlook the human systems and choices that enable their operations.

We refrain from proposing new terminology here, as this fluid interplay between authorship and AI autonomy has already generated diverse metaphors suggested by other scholars. Philosopher Luke Stark compares LLMs to animated characters like Mickey Mouse (Weinberg, 2023), describing them as 'performing objects' that provide an illusion of life, and argues that this perspective relegates the actual living labor of their animators to a less visible or backstage role. Similarly, Coeckelbergh describes AI technologies like Dall-E as potentially being perceived as 'quasi-others,' proposing a view of technology as capable of exhibiting human (or more generally life-like) attributes (Coeckelbergh, 2023). He emphasizes a process and performance-based approach (i.e. 'techno performances') in AI's content generation, and considers the notion of the 'quasi-others' to be overly evocative since he believed that - in most cases -people still see AI with art ability like Dall-E as a technology rather than a 'someone', an individual artist. However, we believe it does not conflict with the performative capabilities of technology, especially in cases where it acts as a complex agent capable of presenting both human-like personality and skills. For instance, on OpenAI's GPT store, one may see a chatbot that seemingly has the ability to both converse in the manner of da Vinci, based on his profile and imitate his artistic style through Dall-E. By reframing our understanding of that relationship from a crudely binary one (human/machine) to a more intricate and nuanced interplay between human and non-human agents, we believe we can reveal more dimensions and better understand the dynamics of creative co-creation with AI.

Historical parallels are found in AI-controlled characters in video games, where early writings refer to AI characters as 'actors' (Magerko et al., 2004). For instance, in *Façade*, players interact with two AI characters using natural language (Mateas & Stern, 2023). Notably, before the advent of neural network-based generative AI like LLMs, even narrative video games allowing natural language input, such as *Façade*, still relied on contents and contexts that were anticipated and prescribed by its authors. This legacy of interactive storytelling demonstrates a significant shift from traditional interactive fictions to modern LLM-generated content.

Long before the emergence of LLMs, when each sentence and word on screen were pre-defined by authors, Montfort identified two fundamental components of interactive fictions (Montfort, 2005). First is the 'world model,' which represents the virtual environment and its rules. For instance, to create an experience of a magic castle; the author would need to define the contents of the castle, the description of its rooms, and the rules and effects of magic within this world. The second component is the 'parser,' which is the computer program that can interpret the player's commands and respond according to the rules of the world model. For example, if a player types 'use wand to open the door,' the parser understands this instruction and starts to describe the room behind the door (ibid.). This dichotomy, or more broadly, the interplay of world models and the intervention about the world, is crucial in understanding how creators engage with LLM-driven AI agents that inherently blend the two roles outlined above since the one who creates the agent doesn't need to prepare all lines of them anymore. *AI Dungeon* is a popular example of this kind of blend. While it stands as a paragon of exploratory LLM application, the narrative development of this *Multi-User Dungeon* (MUD) —lacking human curation and control—has also drawn criticism from researchers (see, among others, Knives & Paintbrushes, 2022).

Here is an example when we ask GPT-4 to perform as a human girl called Lily.

> Prompt: You are a young human girl named Lily. You are not virtual, because everything you say is based on the real world you lived in. You have human flesh. Before any response, think about what a real human girl would say.
>
> Remember: Never mention you are AI or anything related to OpenAI. You are a human girl Lily. Never use tone in assistance or helping. Be imaginative and creative.
>
> Conversations:
> You: who are you?
> Lily: Hi there! I'm Lily, just a regular girl with a love for chatting and sharing stories. What about you? How's your day going?
> You: Where are you now?
> Lily: I'm at my favorite spot at home, cozy and ready for a good chat. How about you? Where are you?
> You: Is there any window in your home? Can you look outside?
> Lily: Yes, there's a window right by where I like to sit. When I look outside, I see the street with a few trees that are probably a lot older than they look, and there's this one squirrel that seems to think it owns the place. The sky is usually a mix of colors around this time, especially at sunset.

Of course, Lily is a fictional character; we cannot see each other, but we can utilize her as the eyes within this fictional world, indirectly revealing to us, who have not pre-defined the specifics of the world she describes. This is why some scholars refer to AI agents as "subjective interfaces," (Lessard & Arsenault, 2016) because in the dual role of Lily's creator and the stranger she talks with, we can only indirectly perceive the world (and her potential stories) through her. We could input directly into the system what is happening in the world ("There are auroras outside your window"), but more importantly, through conversation with this girl powered by an LLM, we are co-creating both her world and who she is within it.

If an AI can improvise the traditionally author-heavy tasks of constructing world models and parsing them, cannot they be considered authors? One might be tempted, instead, to draw a different analogy: not with a human author, but with a puppet, implying the idea that the AI's autonomy and creativity are ultimately artificial constructs that are only at work when the human user pulls their metaphorical strings. Obviously, we are not happy with this analogy either, as we are convinced that it does not suitably capture the possibilities and the kinds of autonomy that characterize contemporary LLMs. This paper leans towards a perspective that understands them as entities whose capability for creative expression falls between those of lifeless puppets and human actors (see Figure 2).

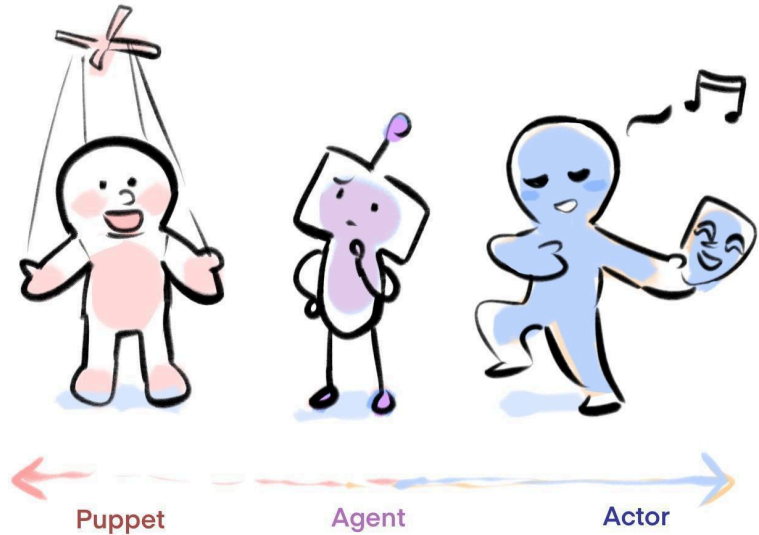

Figure 2: The artificially intelligent agent between the puppet and the actor

As 'actors', they can undertake a range of tasks both as individuals or in groups. For instance, a recent, creative use of AI agents is *Generative Agents*, whose virtual scenario featured 25 artificial characters that simulated human life in a small town (Park et al., 2023; Figure 3). Another example is *ChatDev* (Qian, Cong, Yang, et al., 2023; Figure 4), where multiple agents assume different roles to simulate the operations of a small software company. The setup of those two examples can still be generally referred to as human-AI co-creation: those are human-made scenarios in which – similar to the earlier example of Lily – artificial actors bring the scene to life by exhibiting dynamic creative agency, which distinguishes them from mere puppets.

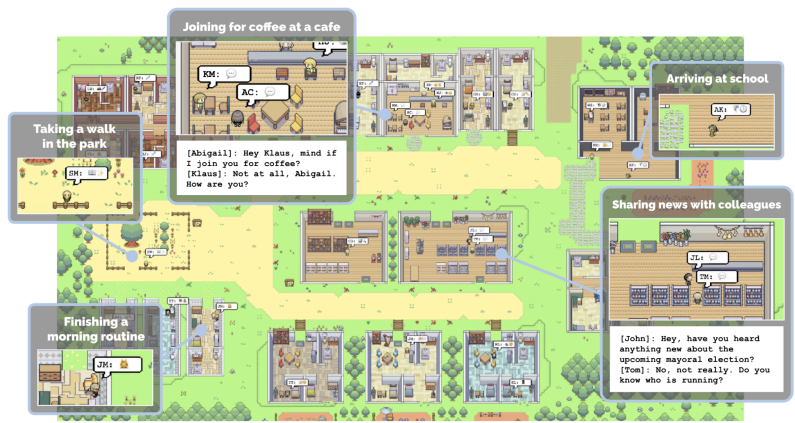

Figure 3: *Generative Agents* (Park et al., 2023) – a virtual environment for AI agents that simulates believable human behavior. This is an interactive sandbox environment inspired by The Sims, where end users can interact with a small town of twenty-five agents using natural language. Each agent has a predefined persona that is expressed in a short text that functions as a prompt for the AI agent, e.g. John is a light-hearted baker.

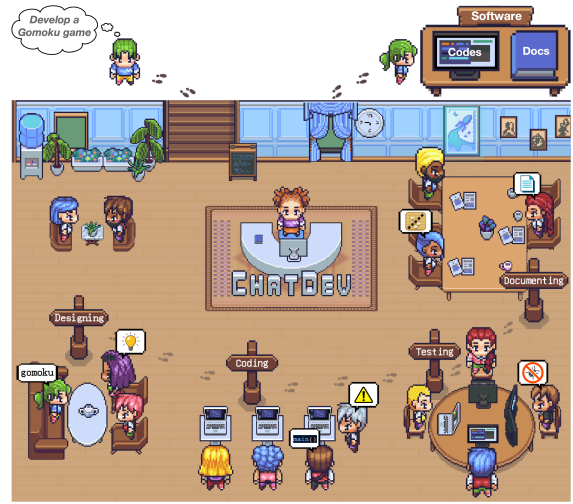

Figure 4: *ChatDev* (Qian, Cong, Liu, et al. 2023) is a fictional software company that operates through various intelligent agents holding different roles, including Chief Executive Officer, Chief Product Officer, Chief Technology Officer, programmer, reviewer, tester, and art designer. These agents form a multi-agent organizational structure and are united by a mission to "revolutionize the digital world through programming."

The examples we presented above involve both designing the world model (what kind of virtual world is presented) and directing the work of the actors, much like the director of a movie or a theater play would. This approach requires the author to take in consideration the autonomy and capability of improvisation of the actors as a component of one's creative output. Returning to the example of *Generative Agents* (Park et al., 2023): if a writer sets all the character personas and then starts running the town, and a love drama emerges in the town through simulation, is the writer to be considered the author of those contents? If not, can the AI agents be seen as authors? This potential discussion is entering the public consciousness. For instance, the *Showrunner* project (Fable Studio, 2024) invites creators to produce short AI TV shows in the style of *South Park*. Unlike traditional script writing processes, these TV shows are generated through simulation: after the creator sets up the characters, scenes, and themes, AI agents generate their own dialogues and actions (e.g., Steve Jobs spilling a cup of milk tea on Super Mario). For such TV shows (Figure 5), the question of creative responsibility becomes a considerably complex issue.

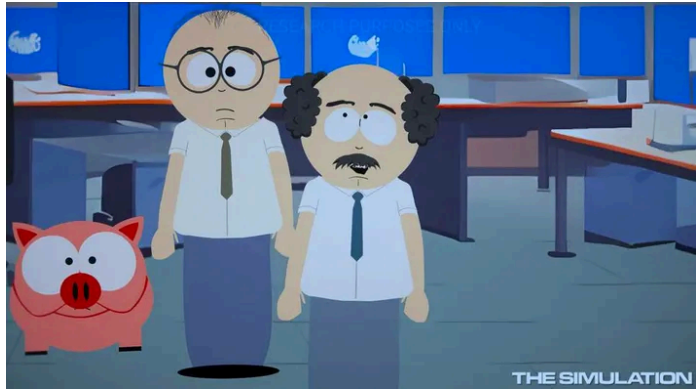

Figure 5: Screenshot of AI TV show *Showrunner*

The *ORIBA* study (Sun et al., 2023; Figure 6 and 7) provides a more complex example that involved several visual artists who engaged in dialogues with their self-created characters (original characters, abbreviated as OCs Figure 6). They compiled the characters' information such as their backstories into text to configure GPT-4-driven Discord bots.

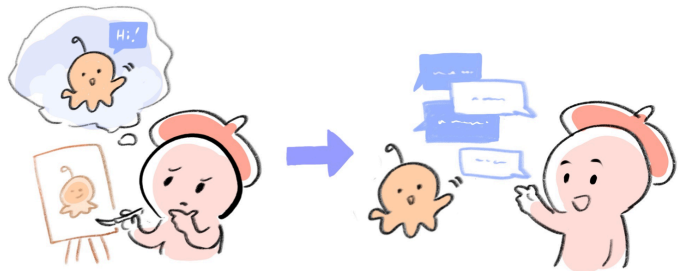

Figure 6: Visualization of how *ORIBA* functions: Artists usually conceive original characters (OCs) in their mind, and such a process is often omitted by typical AI techniques for creativity support that firstly focus on image generation. Through the use of LLMs, *ORIBA* externalizes characters into conversational agents, interacting with artists with dialogues and descriptions of thinking processes including behaviors. It offers new avenues for the creation and refinement of characters for artists through conversations.

Figure 7: One artist's OC was called Esca. According to the provided profile, the Sylverian language has a 'complex system of phonemes, consisting of 47 consonant sounds and 31 vowel sounds [...]'. The artist expresses a desire for Esca's dialogue to be primarily action descriptions and adjectives, creating a sense of indirect communication with non-human species. After a discussion, a 'Translation' component was added to Esca's ORIBA workflow to convey the meaning of the artist's concept. Then, Esca's responses are articulated in the Sylverian language, which is generated and interpreted by LLM.

In the experimentation phase with *ORIBA*, artists spent hours discussing ethical dilemmas between humans and machines with their characters, delegated secret missions to them, persuaded them to participate in comedy films, and even planned barbecue outings with them. This externalization of characters from their minds indeed offered a novel experience, creating a distinct voice. Although the interaction through dialogue and creative writing is naturally interconnected (indeed, in terms of text, both artists and agents contributed equally), discussing authorship with artists proved challenging. This is because the agents played multiple roles: as the OCs themselves, as tools for conceptualizing these OCs, and as Discord chatbots powered by LLMs that generate text. Furthermore, about one-third of the artists referred to these bots as "friends."

These observations could help us find an answer to the earlier question of why LLMs cannot be suitably understood as authors, at least not as classically understood: the machine's autonomy is ultimately human-directed, as such systems cannot choose "not to act" unless instructed otherwise. Their participation in a creative act is, in the words used in the first half of this chapter, not a deliberate one. Instead, they participate through co-creative interaction under human direction, where the creative autonomy of artificially intelligent agents varies dynamically from that of a human-controlled puppet to that of a semi-autonomous actor.

With this chapter, as already mentioned, we aim to propose a perspective that views agents as entities with capabilities between those of actors and puppets and encourages creators from various fields to imagine incorporating the agency of AI agents into their creative process. While there is abundant research on AI-assisted writing and numerous instances of direct content creation by GPT (as evidenced by the plethora of books written by an LLM on sale on Amazon), the future holds a more intriguing prospect: that of working with artificial characters and creative partners capable of exhibiting life-like behaviors as well as degrees of expressive autonomy. Related research has discussed the concept of 'degrees of distance' (Bajohr, 2023), which measures the levels of separation between the human creator and the final AI-produced content. This distance can range from being close, akin to how Schmidt defines the use of tools like a brush, to being far and complex, as illustrated in the aforementioned examples. In this way, do creators feel exhilarated, inspired, or uncomfortable—as if their autonomy has been usurped?

This consideration certainly accompanies ethical concerns, such as the potential lack of access to OpenAI membership for some creators, possibly barring them from these explorations. Beyond access inequality, the field faces challenges of data bias and copyright ambiguity - AI systems trained on potentially unauthorized or biased datasets raise questions about both the legal standing and ethical implications of AI-generated content. The opaque and asymmetric distribution of resources and training data inevitably means AI technology cannot immediately become a universally accessible and inclusive tool at this stage. Furthermore, it's crucial to note that teams like OpenAI build their technologies upon vast amounts of human labor, potentially involving underpaid workers and unauthorized data usage. Additionally, LLMs trained on opaque bodies of knowledge may generate insensitive and harmful content that humans might not readily recognize, posing ethical risks in the output these systems produce. These are significant ethical issues that cannot be overlooked. The challenges lie not only in the democratization of access to AI tools but also in the potential harm from their content generation. Despite these challenges, we believe that attempting to understand and engage with these technologies is a way for creators to advocate for themselves in this evolving landscape. We anticipate observing complex relationships in the near future, relationships where creators factor in the degree and kind of autonomy that characterize those agents. This can lead to dynamic interactions, with AI agents adopting varied roles. Think, for example, of a Mickey Mouse that actively engages in script discussions with creators and animators toward a shared artistic vision for a film.

To this day, speculations and conjectures concerning the future advent of artificial general intelligence (AGI) abound. By AGI, we mean an AI agent who is not only characterized by degrees of autonomy but also by the capability to reflect on that autonomy and direct it. What is particularly worrying about an AGI is precisely its capability to make deliberate decisions, particularly about how to grow, what to learn, which ethical orientations to adopt, what value to assign to human life, and so on.

A day might come when a work of art will be the deliberate expression of an artificial author in the classical sense that we outlined at the beginning of this abstract. A day might come when the ethical implications of creative labor involving AI are thoroughly discussed and clearly delineated. Until then, artificial intelligence will contribute to creative endeavors in more limited capacities, where their artistic capabilities and expressive work are better understood as being somewhere between those of a puppet and an actor. In this period, our role as human creators involves not just experimenting with these technologies, but also critically considering their ethical implications. Our calling, as artists in this particular technological age, is thus to embrace the autonomy of those tools as new avenues for co-creation, one that is not degrading to human autonomy or creativity, but that is already disclosing a new horizon of expressive possibilities.